\documentclass[epj,twocolumn]{webofc}
\usepackage {amsmath}
\usepackage[varg]{txfonts}   
\woctitle{Web of Conferences}
\begin{document}
\title{New Cosmological Model and Its Implications on Observational Data

Interpretation}

\author{B. Vlahovic\inst{1}\fnsep\thanks{\email{vlahovic@nccu.edu}}
}

\institute{North Carolina Central University, Durham, NC, USA
          }
\abstract{%
  The paradigm of $\Lambda CDM$ cosmology works impressively well and with the concept of inflation it explains the universe after the time of decoupling. However there are still a few concerns; after much effort there is no detection of dark matter and there are significant problems in the theoretical description of dark energy. We will consider a variant of the cosmological spherical shell model, within FRW formalism and will compare it with the standard $\Lambda CDM$ model. We will show that our new topological model satisfies cosmological principles and is consistent with all observable data, but that it may require new interpretation for some data. Considered will be constraints imposed on the model, as for instance the range for the size and allowed thickness of the shell, by the supernovae luminosity distance and CMB data. In this model propagation of the light is confined along the shell, which has as a consequence that observed CMB originated from one point or a limited space region. It allows to interpret the uniformity of the CMB without inflation scenario. In addition this removes any constraints on the uniformity of the universe at the early stage and opens a possibility that the universe was not uniform and that creation of galaxies and large structures is due to the inhomogeneities that originated in the Big Bang.

}
\maketitle
\section{Introduction}
\label{sec-1}
The $\Lambda CDM$ cosmology in combination with the inflation model gives the best predictions for the observable data. For instance GR and standard model can predict with high accuracy decreases in the orbital period of a binary pulsar and angular power spectrum of the CMB. Inflation can explain horizon, flatness, and monopole puzzles and give a natural quantum mechanical mechanism for the origin of the cosmological fluctuations observed  in the CMB and in the large scale structure of matter. However, there are not observable/experimental tests to give us any assurance that an inflation epoch really left measurable effects. The explanation for flatness may be the anthropic principle \cite{Barrow}, that intelligent life would only arise in those patches of universe with $\Omega$ very close to 1; another explanation could be that space is precisely flat, so that $K=0$ now and always. Guth's monopoles may be explained by inflation, or the physics may be such that they never existed in appreciable abundances. An explanation may be that there is no simple gauge group that is spontaneously broken to the gauge group $SU(3)\times SU(2)\times U(1)$ of
 the Standard Model.

This paper is not about the criticism of inflation models, such as that: it invokes an inflation field that does not correspond to any known physical field, that its potential energy curve is an ad hoc contrivance to accommodate observable data,  and that inflation does not solve the problem of initial conditions, because it requires extremely specific initial conditions of its own. In the new inflation model there are also fine tuning requirements that the universe must have a scalar field with an especially flat potential (compared to the large vacuum energy), small first and second derivatives, and that the inflation particles must have a small mass.

Since there are possible solutions of the flatness and the monopole problems that do not rely on inflation, we will focus on the horizon or homogeneity puzzle, for which there are no proposed reasonable definite alternative resolution.  We will also show that our proposed solution of the horizon problem can also give a reasonable explanation for the formation of the large scale structure of matter.
Actually the classical model does say that expansion traces back to a singular state, but the observed large-scale homogeneity and isotropy is not required by classical GR theory.  It is well known that in the Big Bang models homogeneity of space cannot be explained, it is simply assumed in initial conditions. Homogeneity in the CMB on the level of $10^{-5}$ is explained by inflation era. However, arguments in favor of inflation only exist if space was already homogeneous before inflation. If the pre inflationary universe was not already homogenous, inflation will not lead to homogeneity \cite{Goldwirth}.

 So, the homogeneity problem is pushed only back in time, because the Big Bang itself is taken to be inherently free of correlations.
We argue that the observed uniformity in the CMB does not mean that space was uniform at the time of decoupling. We propose a shell cosmological model that allows for a different interpretation of the CMB data and for inhomogeneity of the universe at the early stage.

The uniformity of the CMB is not consistent with the standard model without inflation. The curvature and physical properties of the regions of space which have never been in causal contact and should not be correlated are taken to be indistinguishable. During the matter dominated era the scaling factor $a(t)$ increases as $t^{2/3}$, since the time of last scattering the horizon size was of the order
  $d_H  \approx H_0^{-1} (1+z_L)^{-3/2}$, where $z_L$ is redshift at the surface of last scatter. The angular diameter distance $d_A$ to the surface of last scattering is of order $H_0^{-1} (1+z_L)^{-1}$, so the horizon at the time of last scatter now subtends an angle of order $d_H/d_A  \approx (1+z_L)^{-1/2}$, which is for $z_L \simeq 1100$ about $1.7^{\circ}$. Therefore in CMB spectrum points further apart than several degrees should be not correlated, but correlations up to $\sim 60^{\circ}$ are observed.

\begin{figure}
\includegraphics[width=8cm]{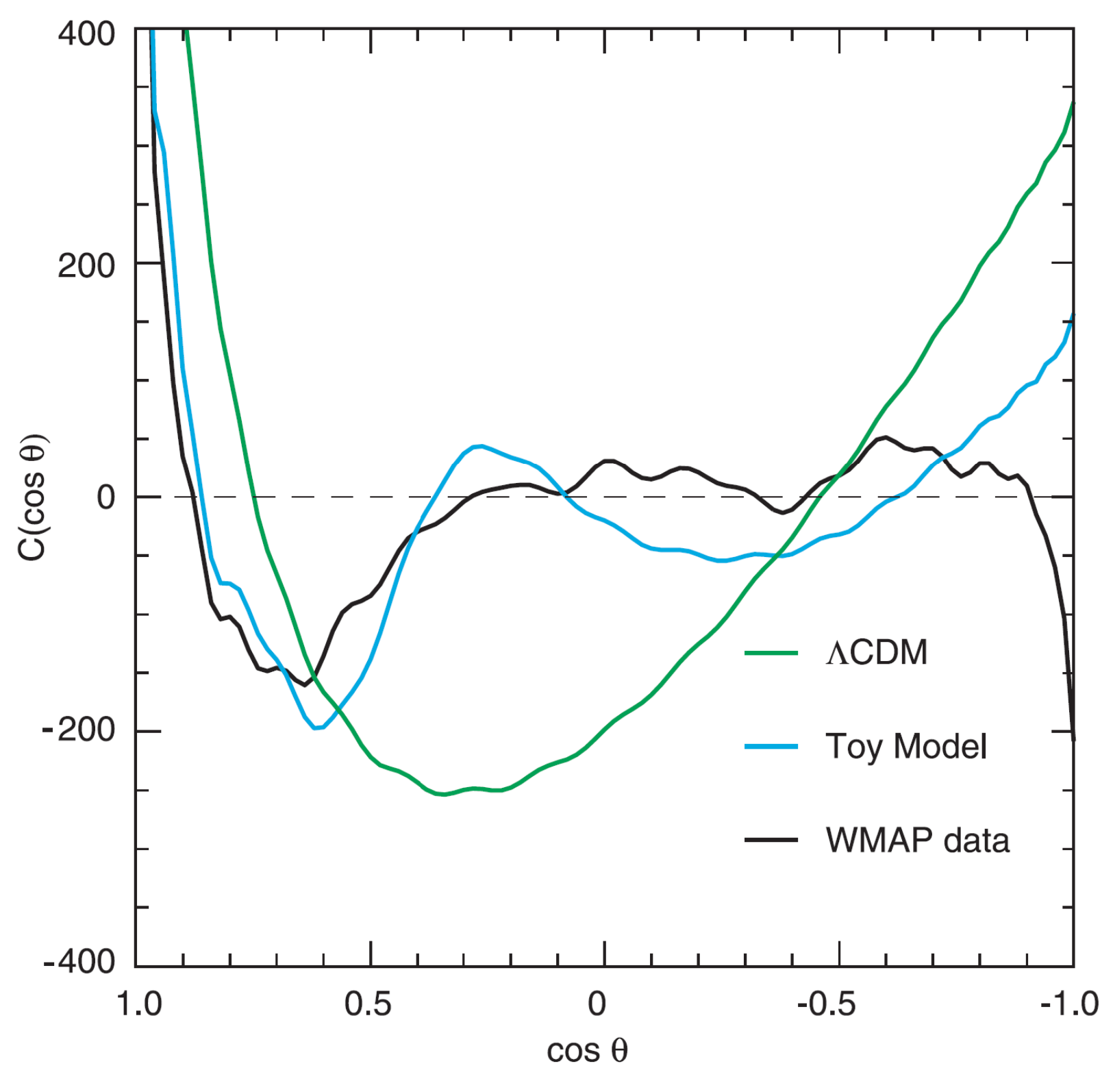}
\caption{\label{CMBCorrelations}
Angular correlation function of the best fit $\Lambda CDM$ model, a finite size universe model, and
WMAP data on large angular scales (adopted from \cite{CMBWMAP1}).}
\end{figure}

\begin{figure}
\includegraphics[width=8cm]{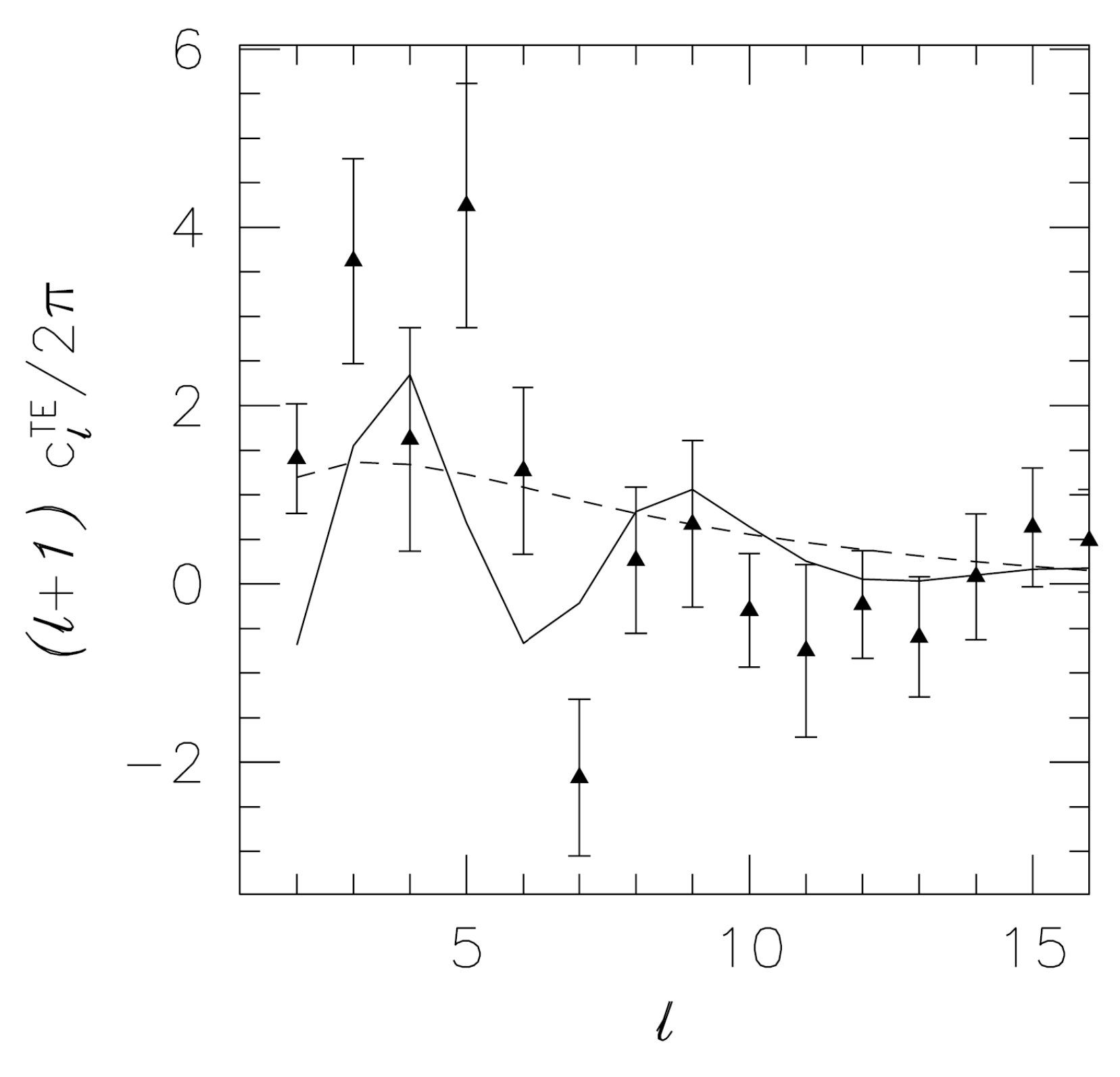}
\caption{\label{CMBTE}
The comparison of the data to the predicted TE power spectrum in a finite universe model (solid line) and the $\Lambda CDM$ model (dashed line), adopted from \cite{CMBWMAP1}.}
\end{figure}

Inflation is also not consistent with observed large scale angular correlations in CMB data. Inflation models require angular correlation at all angles, not only at angles up to $\sim 60^{\circ}$, because inflation occurred at all scales.  The discrepancy in angular correlations between CMB and $\Lambda CDM$ model that is presented in Fig. \ref{CMBCorrelations} was first noticed in \cite{CMBCOBE} and confirmed later in \cite{CMBWMAP1}.
   There is an obvious difference between the CMB spectrum and predictions of the standard model. The figure also includes a curve that shows very good agreement between the observable data and a finite size universe model (similar to the model proposed here). Please note that the finite size model gives not only a better match to the observed correlation function than the  $\Lambda CDM$ model, but also predicts the distinctive signature in the temperature polariaztion $(TE)$ spectrum; see Fig. \ref{CMBTE}.


\section{Shell Model and uniformity of CMB}
\label{sec-2}

GR does not specify the topology of space. Einstein's equations describe only local properties of the spacetime, but do not fix the global structure and topology of spacetime. Different topologies can correspond to the same matrix element, leaving the possibility of new models of the universe.

We consider a model in which the universe is an expanding spherical shell with a significant thickness, \cite{varxiv}. The size of the shell (the arc length from pole to antipode) must correspond to the present size of the cosmological horizon, and the thickness of the shell must have a minimum size to explain present observation constraints: ghost images of sources; distribution and periodicity of clusters, super clusters, quasars, and gamma-ray bursts; statistical analysis of reciprocal distances between celestial objects; and other limits obtained from the CMB (uniformity and weak angular fluctuations). For example, from the statistical analysis of the Abell catalog of spatial separation of clusters, it appears that the shell thickness should be at least about 1 Gpc \cite{Lehoucq}.

 Because there is no other space than that associated with the spherical shell the motions of all galaxies and propagation of the light must be confined to the volume of the shell, which expands with a radial velocity. As we will see this will have significant implications on the interpretation of the data. The light must follow geodesic lines, so as in torus models it is not traveling straight, but is bent.

\begin{figure}
\includegraphics[width=8cm]{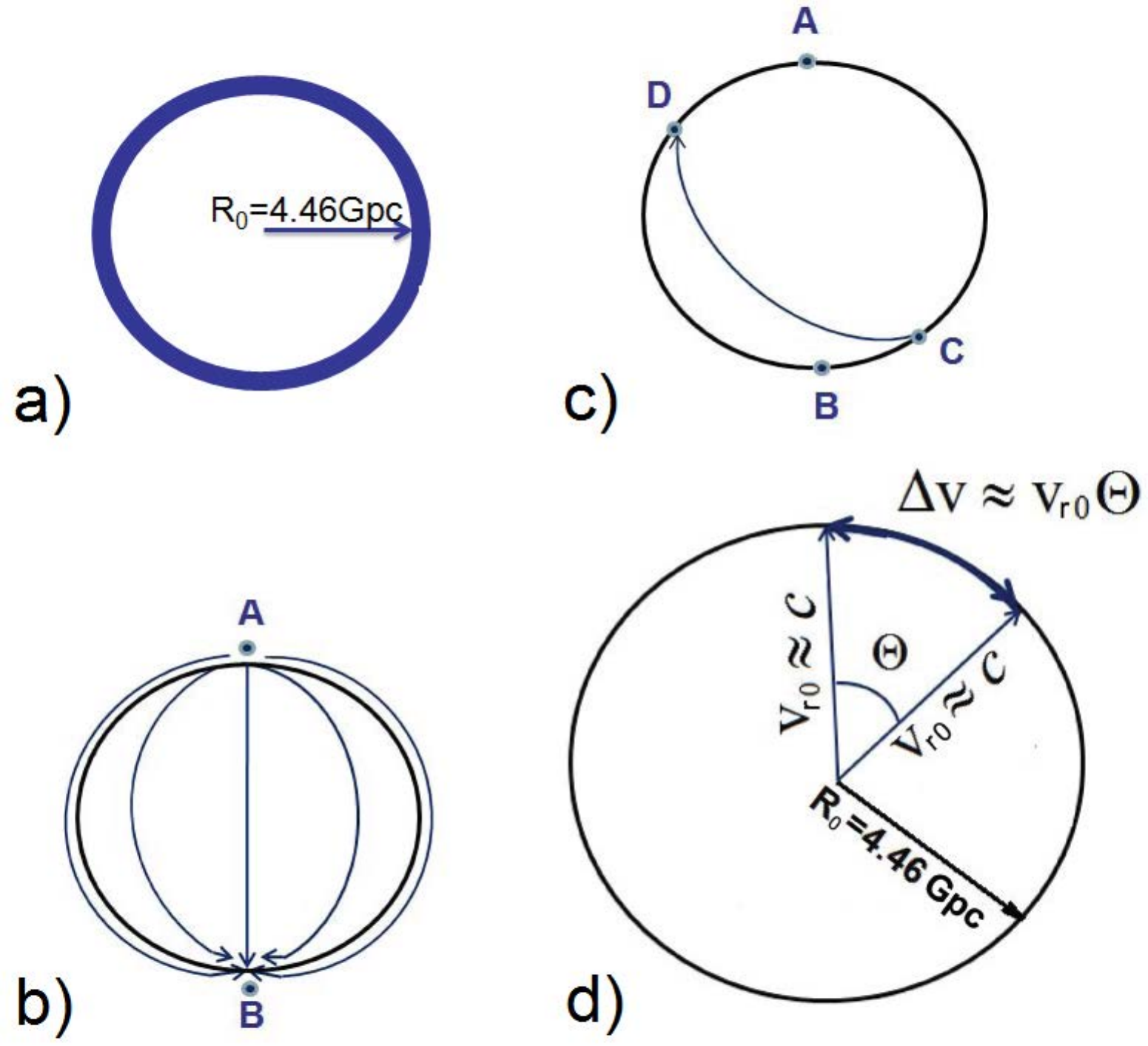}
\caption{\label{Model}
a) The visible universe as an expanding shell with a significant thickness, b) The observable universe, as seen by an observer from point A, is a volume of the shell, with event horizon located at point B, c) CMB visible from Earth (by observer at point A) originates at the antipodal point B and CMB visible from another place in the universe (point C) is emitted at point D, d) The visible universe as a surface of the sphere with radius $R_0$ = 4.46 Gpc that expands with radial speed close to the speed of light.
}
\end{figure}

 This will define an observable universe for our model as the largest visible volume inside the spherical shell, from the point of the observer, Fig. \ref{Model}a. By this definition the cosmological horizon distance will be the largest possible arc distance along the spherical shell. If the universe is the same size as the observable universe and an observer is located at point A, the particle horizon distance will be the arc distance from point A to point B located at the antipode and the observable universe will be the entire spherical shell.

The dynamics of shell models have been investigated earlier. For a systematic study in the framework of general relativity, see for instance \cite{Berezin} and \cite{Krisch}. However, this is not the focus of our paper. Regardless of the model that will be used, a simple expansion of the shell with a constant speed introduced in \cite{Vlahovic}, a GR model first introduced by Israel \cite{Israel}, or the special relativity model \cite{Czachor}, the agreement with the observable data will be nearly the same. In addition, because the propagation of light is confined to along the shell (for instance it cannot travel across the shell, because there is no space) all these models will require a new interpretation of the observable data, for instance uniformity of the CMB, as it will be shown here.

In the current $\Lambda$CDM model the visible universe is defined as a sphere centered on the observer and from our perspective it appears that the radius is $R_0$ = 14.0 $\pm$ 0.2 Gpc (about 45.7 Gly). The value $R_0$ is the particle horizon and the quoted result corresponds to the direct WMAP7 measurements and the recombination redshift $z$ = 1090 $\pm$ 1 \cite{2}.

In the standard FLRW model
\begin{equation} \label{radius__1}
R_o = a(t) \int_{t'=0}^{t}\frac{c}{a(t')}dt',
\end{equation}
where \begin{equation} \label{radius__2}
\frac{da}{dt} = \sqrt{\frac{\Omega_r}{a^2} + \frac{\Omega_m}{a} + \frac{\Omega_\Lambda}{a^{-2}}}.
\end{equation}
The $R_0$ = 14.0 $\pm$ 0.2 Gpc corresponds to the following combinations of the parameters: $\Omega_m h^2 =0.136 \pm 0.003$, $\Omega_r = \dfrac{8\pi G}{3H^2} \dfrac{\pi^2 k^4 T^4}{15 c^5\hbar^4}$, $\Omega_{\Lambda} = 1 - \Omega_r - \Omega_m$.

In our shell model we must obtain the same value for the particle horizon, which is now an arc distance from pole to antipode, instead of the radius of the sphere. For closed universe in FRW metric
\begin{equation} \label{metric}
ds^2= c^2dt^2-a(t) ^2 [d\chi^2 + \sin^2\chi (d\theta^2 + \sin^2\theta d\phi^2)],
\end{equation}
where $\chi=\int cdt/a(t)$, which is by definition the conformal time coordinate or arc parameter time $\eta$. During the interval of time $dt$,  a photon traveling on a hypersphere of radius $a(t)$ covers an arc $d\eta = dt/a(t)$.
The luminosity distance $d_L$ is related to the radial comoving coordinate
\begin{equation} \label{dL}
d_L= r\sin(r/R_0)(1+z),
\end{equation}
where $\sin(r/R_0)$ reflects the positive curvature and that the photons are spread over a smaller area $A_p(t_0)$ than they would be in the flat space, $A_p(t_0) = 4\pi r^2\sin^2(r/R_0) < 4\pi r^2$.

The luminosity distance depends upon the expansion history through $\int cdt/a(t)$ and the curvature through $\sin(r/R_0)$. For the simple model that we consider, the shell expands with constant speed $c= H(z)R(z)$
\begin{equation} \label{dLH}
\begin{split}
d_L& =r\sin(r/R_0)(1+z) =  R_0(1+z)\int_{0}^{z} \frac{\sin(r/R_0)}{H(z')}dz' \nonumber \\
   & =\frac{c}{H_0}(1+z)\int_{0}^{z} \frac{\sin(\pi / [2(1+z/\pi)])}{1+z}dz'.
\end{split}
\end{equation}
\begin{figure}
\includegraphics[width=8cm]{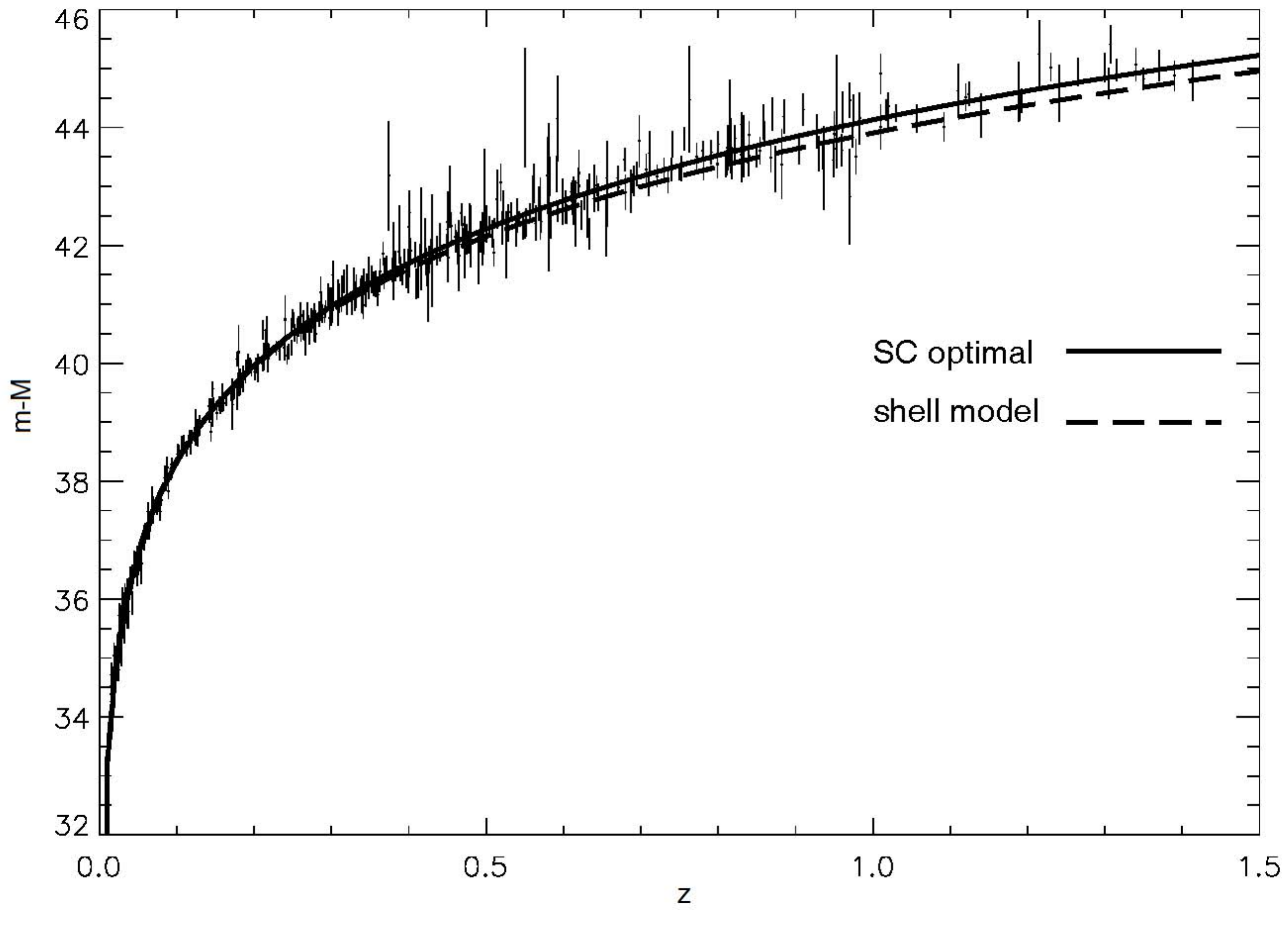}
\caption{\label{Modulus}
Hubble diagram for the Union2.1 data set \cite{Suzuki 2012}. The solid line represents the best-fitted $\Lambda CDM$ model. The dashed line represents the shell model, with the shell  expanding with speed $c$.}
\end{figure}
In Fig. \ref{Modulus} we compare the distance modulus for the simple shell model expanding with constant speed, obtained through equation (\ref{dLH}) dashed line, with the distance modulus for the standard $\Lambda CDM$ model obtained by
\begin{equation} \label{dLHLCMD}
d_L=\frac{c}{H_0}(1+z) \int_{1/(1+z)}^{1} \frac{dx}{\sqrt{\Omega_r + x\Omega_m + x^4\Omega_{\Lambda}}},
\end{equation} where $x=a/a_0$.
\begin{figure}
\center
\includegraphics[width=8cm,height=5.0cm]{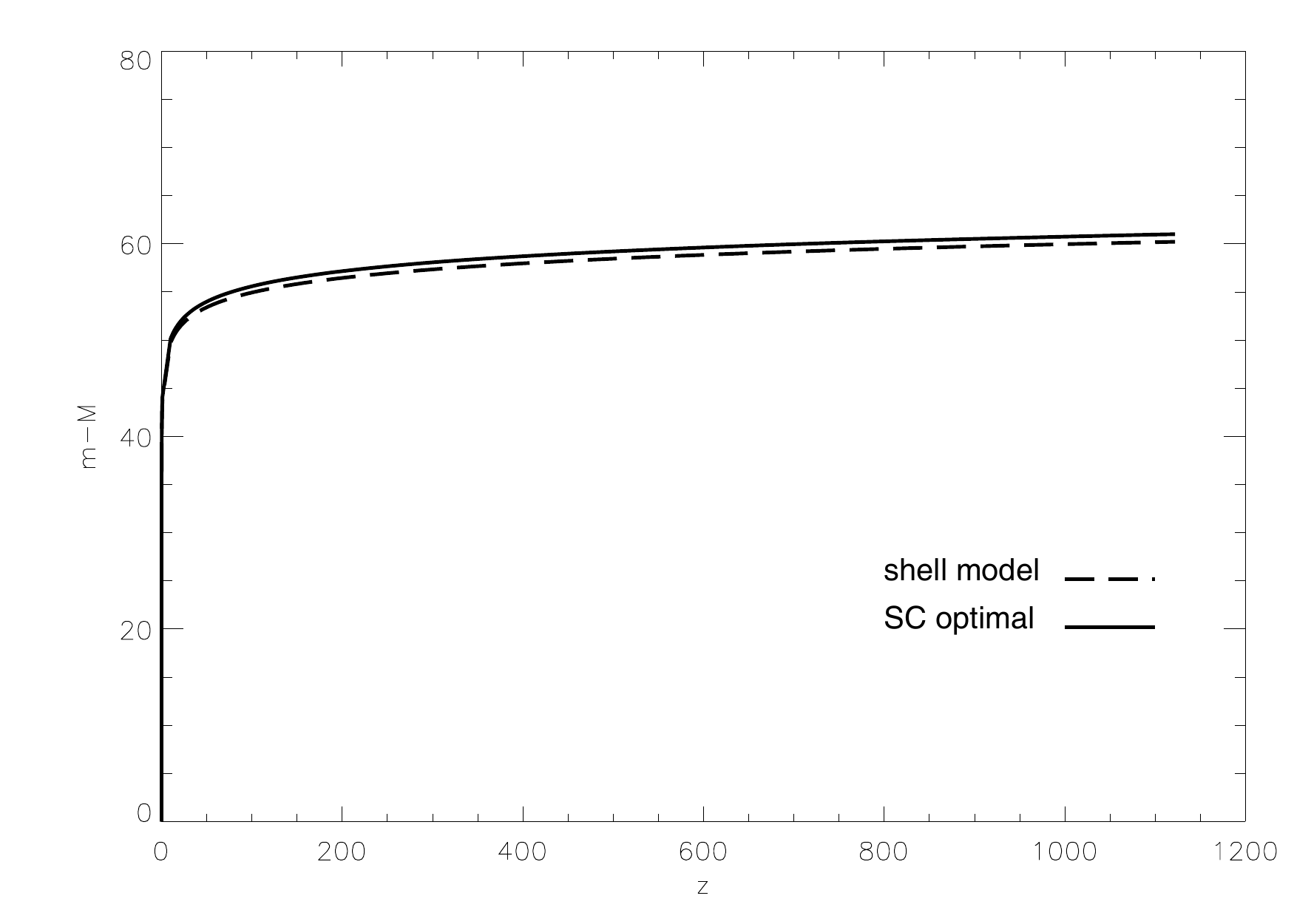}
\caption{\label{figtwo_h_67_26_1000}
Comparison of the distance modulus for the $\Lambda CDM$ model (solid line) and the shell model expanding with speed $c$ (dashed line).}
\end{figure}
Fig. \ref{figtwo_h_67_26_1000} compares the  standard $\Lambda CDM$ and the shell model considered here. The $\chi^2$ statistic for the spherical shell model is also calculated and compared with the best fit cosmology
\begin{equation} \label{Chisquare}
\chi^2 = \sum_{i}   \frac{[\mu_i - 5log10(d_L(z_i)/10pc)]^2}{\sigma^2_i}
\end{equation}
where $\mu_i$ is distance modulus for a detected supernova with index $i$, $d_L$ is luminosity distance,
and $\sigma^2_{i}$ is dispersion for $\mu_i$ evaluation.
The $\chi^2=563.81$ for the $\Lambda CDM$ and $\chi^2=734.19$  for the simple shell model that expands with speed c. It will be interesting to see how much $\chi^2$ may improve for the shell model, if the GR shell model is used. Additional improvement may also be possible through corrections due to the gravitational lensing, which in this model could be important, and by recalculating the data set using "nuisance" parameters that correspond to the shell model. However, even this result indicates that the proposed shell model is in concordance with Union2.1 catalog data.

The first consequence of this model is that the curvature radius of our spherical shell will be a factor of $\pi$ smaller than in the standard FLRW model, see  Fig. \ref{Model}d and \cite{varxiv}. This will result in significantly different density, which will be at least a factor of $\pi^3$ (depending on the thickness of the shell) larger in the spherical shell model.


The most important consequence of this model is that the confinement of the light along the shell leads to the observed uniformity of the CMB.  Because regardless of the direction we chose to measure CMB (for instance from point A looking in any direction), we will always measure CMB at the antipodal point B.

Therefore measuring the same CMB by looking in the opposite directions of the universe does not represent or reflect the uniformity of the universe at the time of decoupling, because we always measure CMB originating from the same point regardless of the direction of observation. For that reason we always must obtain the same result. If from point A we observe in any direction, we will always measure the CMB originating from point B. Small variations for the CMB are possible and observed, but these variations are the result of the interaction between matter and light during its travel. For instance, depending on the direction we choose to measure the CMB, light will travel from point B to A through different galaxies and will interact with different amounts of matter, which will result in the small observed variations of the CMB. The observed fluctuations in the CMB are therefore created as the photons pass through nearby large scale structures by the Integrated Sachs -- Wolfe effect \cite{4a}\cite{4b}\cite{4c}\cite{4d}.

To establish a connection between the uniformity of the earlier universe at the time of decoupling and the CMB we will need to make a completely different kind of measurement of the CMB. We can see the CMB in any direction we can look in the sky. However, we must keep in mind that the CMB emitted by matter that would ultimately form, for instance the Milky Way, is long gone. It left our part of the universe at the speed of light billions of years ago and now forms the CMB for observers in remote parts of the universe, for an observer located at the antipodal point B. For instance, if we perform a measurement of the CMB at the point C, we will measure the CMB emitted by matter at the point D, Fig. \ref{Model}c. To measure the uniformity of the universe at the time of decoupling we will need to measure the CMB in at least two different points on the shell. If, for instance, the measurements from points A (CMB originated at B) and C (CMB originated at D) give the same result, then and only then may we speak about the uniformity of the CMB and uniformity of the universe at the time of decoupling. However, such measurements are not possible at the present time.

\begin{figure}
		\includegraphics[width=8cm]{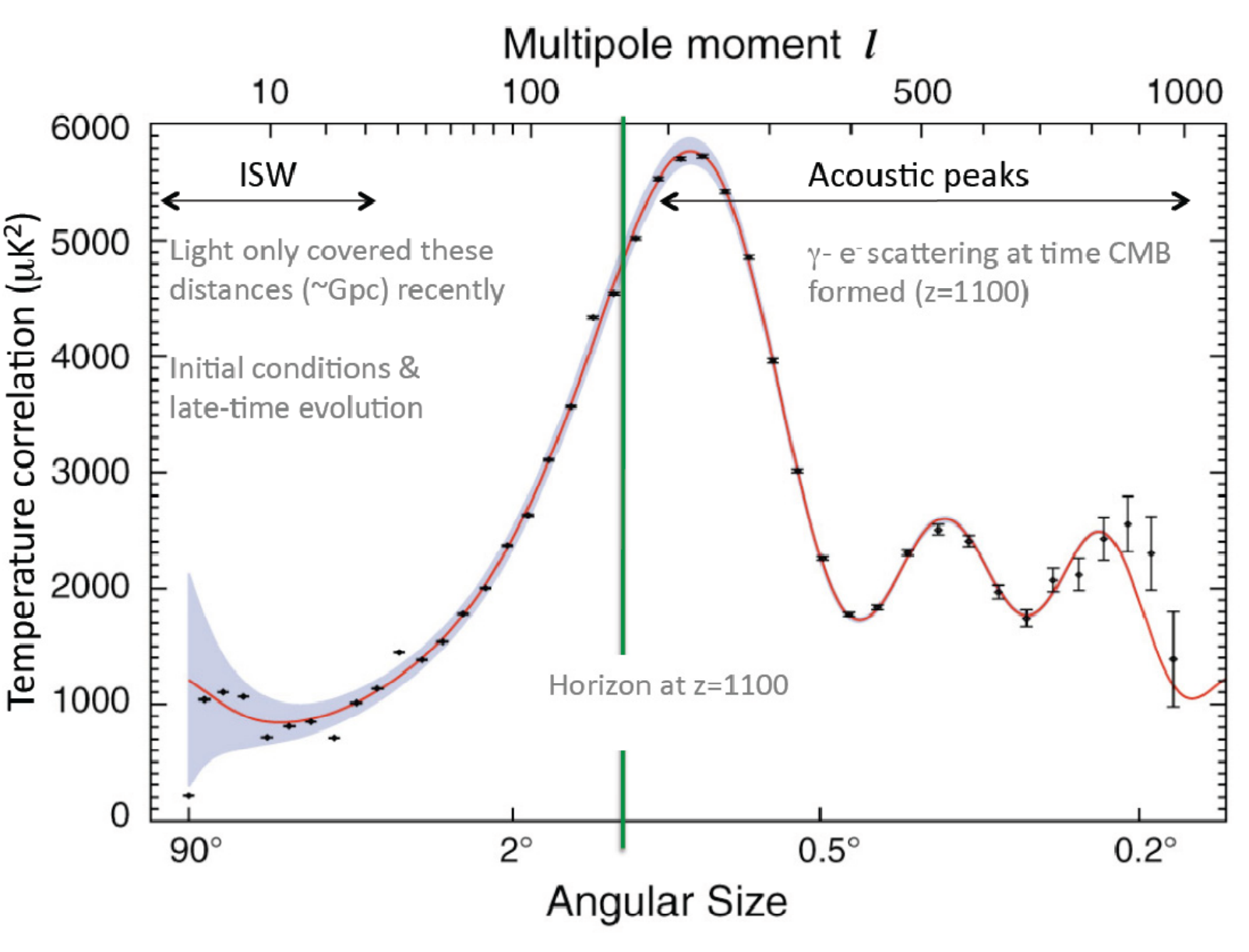}
		\caption{\label{CMBPS}
 The CMB power spectrum as a function of multipole moment $l$ and angular size, adopted from the WMAP collaboration.}
\end{figure}

The CMB temperature fluctuations $\Delta T/T$ are usually written in terms of a multipoles expansion on the celestial sphere:
\begin{equation} \label{deltaT}
                      \frac {{\Delta}T} {T} (\theta,\phi) = \sum_{l=2}^{\infty} \sum_{m=-l}^{l}a_{lm} Y_l^m(\theta,\phi).
\end{equation}
However, what is actually directly measured by observations is the angular correlation of the temperature anisotropy
$\langle \frac {\Delta T} {T} (\hat{n}_1) \frac {\Delta T} {T} (\hat{n}_2) \rangle$ where $\cos\theta = \hat{n}_1 \cdot \hat{n}_2$. This is expressed through the power spectrum $C_l \equiv \langle |a_{lm}|^2 \rangle$, Legendre polynomials, and the filter function $W_l$ as
   \begin{equation} \label{Ctheta}
                      C(\theta) = \frac {1} {4\pi} \sum_{l} \left[ \frac {l + \frac {1} {2}} {l(l+1)} \right] C_lP_l(\cos\theta) W_l.
\end{equation}
        The main contribution to $C_l$ for $l>60$ is from oscillations in the photon-baryon plasma before decoupling, see Fig. \ref{CMBPS}. However, in the spherical shell model we cannot see the imprint of these oscillations in the CMB at the time of last scattering, because we are always measuring CMB coming just from the single point (or space limited region), the antipodal point (region), of the surface of the last scattering. We cannot see the $C_l$ contributions that are from the remaining part of the surface of the last scattering. For instance as we already mentioned, fluctuations from the surface of the last scattering that at the present corresponds to the Milky Way galaxy already left us billions of years ago. Therefore, in principle, we cannot obtain the angular correlation of the CMB temperature anisotropy in the shell model.
        The contribution to $C_l$ for low multipoles $l \leq 60$ is mainly from the Integrated Sachs -- Wolfe effect that relates temperature fluctuations to the integral of variations of the metric evaluated along the line of sight.
One can argue that the line of sight is similar in the 3-sphere and spherical shell models. For instance, assume that we are on the surface of a 3-sphere and that propagation of the light is confined to its surface, then the observed distribution of the galaxies on  the surface of the 3-sphere and in the spherical shell will be the same. We should therefore obtain a very similar spectrum for low multipoles $C_l$  for the spherical shell model and the standard ${\Lambda}CDM$ model.
   Therefore the temperature asymmetry for $C_l \leq 60$ should be similar in both models.

   Another argument for the shell model is that if the universe was finite and smaller than the volume
within the decoupling surface, then there should be a very distinctive signal: matched circles \cite{Cornish},\cite{Cornishb},\cite{Weeks}. These matched circles may be circles detected in \cite{Gurzadyan}, which may be the signature of a finite universe and may help us to determine the topology.
It is also important to note that a hollow shell model completely reproduces the distribution of the entire observed radio sources count for the flux density $S$ from $S\approx 10$ $\mu$Jy to $S\approx 10$ Jy \cite{Condon}.

    \section{Conclusion}

The shell model considered here is in agreement with SNe Ia and other observational data. The model explains the observed uniformity of the CMB without inflation, because propagation of the light is confined along the shell. The entire observed CMB originates from a single antipodal point or small region. The measured CMB must therefore be exactly the same for all directions of observations, if corrected for Integrated Sachs -- Wolfe fluctuations caused by large scale structures. For that reason we cannot say anything about the uniformity of the universe at the early stage. This removes any constraints imposed by the CMB on homogeneity. It is thus possible that the universe was not uniform at an early stage and that creation of galaxies and large structures is due to the inhomogeneities that originated in Big Bang, because there is no reason that the Big Bang should be homogeneous.

{\bf Acknowledgments:}
I would like to thank S. Matinyan, A. Zakharov, and I. Filikhin for useful discussions. This work is supported by NSF award HRD-0833184 and NASA grant NNX09AV07A.


\begin{thebibliography}{}
\bibitem{Barrow} J.D. Barrow, F.J.Tipler, The antropic Cosmological Principle, Oxford, 1986.
\bibitem{Goldwirth} S.D. Goldwirth and T. Piran, in the 6th, Marcel Grossmann meeting on general relativity, H. Sato and T. Nakamura ed., World Scientific, (1992)1211.
\bibitem{CMBCOBE}G. Hinshaw et al., Astrophys. J. Lett. \textbf{464}  (1996)L25.
\bibitem{CMBWMAP1} D. N. Spergel et al., Astrophys. J. Supp. \textbf{148} (2003)175.
\bibitem{varxiv} B. Vlahovic,
Proceedings of Low Dimensional Physics and Gauge Principles, Yerevan, Armenaia, September 21-26 2011, p. 241, and arXiv:1207.1720.
\bibitem{Lehoucq} R. Lehoucq,  M. Lachi\'{e}ze-Rej, and J.P. Luminet, arxiv:gr-qc/9604050 (1996).
\bibitem{Berezin} V.A. Berezin, V.A. Kuzmin, and I.I. Tkachev,
Phys. Rev. D \textbf{36} (1987)2919.
\bibitem{Krisch} J.P. Krisch and E. N. Glass,
Phys. Rev. D \textbf{78} (2008)044003.
\bibitem{Vlahovic} B. Vlahovic, 
arXiv:1005.4387.
\bibitem{Israel} W. Israel, 
Nuovo Cimento \textbf{44B} (1966)1-14.
\bibitem{Czachor} A. Czachor, 
Acta Physica Polonica B \textbf{38} (2007)2673.
\bibitem{2}  S. H. Suyu, et al.,
    The Astrophys. J., \textbf{711} (2010)201.
\bibitem{Suzuki 2012}N. Suzuki,  et al., ApJ, \textbf{746} (2012) 85.
\bibitem{4a} C. Ho, C, Hirata, N. Padmanabhan, U. Seljak, N. Bahcall, 
    Phys. Rev. D \textbf{78} (4) (2008)043519.
\bibitem{4b} T. Giannantonio, et al., 
    Phys. Rev. D \textbf{77} (12) (2008)123520.
\bibitem{4c} A. Raccanelli, et al., 
    Monthly Notices of the Royal Astronomical Society \textbf{386} (4) (2008)2161.
\bibitem{4d} B.R. Granett, M.C. Neyrinck, and I. Szapudi, 
    Astrophys. J. \textbf{683} (2) (2008)L99.
\bibitem{Cornish} N.J. Cornish, D. Spergel, and G. Starkman,  Phys. Rev. D, \textbf{57} (1998)5982.
 \bibitem{Cornishb} N.J. Cornish, D. Spergel, and G. Starkman,  Proc. Nat. Acad. Sci., \textbf{95} (1998)82.
 \bibitem{Weeks}J.R. Weeks, Class. Quant. Grav. \textbf{15} (1998)2599.
\bibitem{Gurzadyan}V.G. Gurzadyan, R. Penrose, 
    arXiv:1011.3706.
\bibitem{Condon} J. J. Condon, 
Galactic and Extragalactic Radio Astronomy, $2^{nd}$ edition, eds. G.L. Verschuur and K. I. Kellermann, 1988.

\end{thebibliography}
\end{document}